\documentclass[aps,prc,preprint,showpacs,superscriptaddress]{revtex4}
\usepackage{epsfig}
\usepackage{amsmath}
\usepackage[hyperref]{hyperref}

\begin{document}

\title{Medium modifications of the bound 
nucleon GPDs and incoherent DVCS on nuclear targets}

\author{V. Guzey}
\email{vguzey@jlab.org}
\affiliation{Theory Center, Thomas Jefferson National Accelerator Facility, 
Newport News, VA 23606, USA}

\author{A.W. Thomas}
\email{awthomas@jlab.org}
\affiliation{Theory Center, Thomas Jefferson National Accelerator Facility, 
Newport News, VA 23606, USA}
\affiliation{College of William and Mary, Williamsburg, VA 23178, USA}

\author{K. Tsushima}
\email{tsushima@jlab.org}
\affiliation{Excited Baryon Analysis Center (EBAC) and Theory Center,
Thomas Jefferson National Accelerator Facility, 
Newport News, VA 23606, USA}

\preprint{JLAB-THY-08-837}

\begin{abstract}
We study incoherent DVCS on $^4$He in the 
$^4{\rm He}(e,e^{\prime}\gamma p)X$ 
reaction, which probes possible medium-modifications of the bound 
nucleon GPDs and elastic form factors.
Assuming that the bound nucleon GPDs are modified in proportion to the
corresponding bound nucleon elastic form factors, as predicted in the 
Quark-Meson Coupling model, we develop an approach to calculate various 
incoherent nuclear DVCS observables. As an example, 
we compute the beam-spin DVCS asymmetry, and 
predict the $x_B$- and $t$-dependence of the ratio of the 
bound to free proton asymmetries, $A_{\rm LU}^{p^{\ast}}(\phi)/A_{\rm LU}^{p}(\phi)$. 
We find that the deviation of $A_{\rm LU}^{p^{\ast}}(\phi)/A_{\rm LU}^{p}(\phi)$ from unity is as much as $\sim$ 6\%.
\end{abstract}

\maketitle

%Intro

Properties of hadrons in a nuclear medium are expected to be modified 
compared to those in the vacuum.
As indicated by measurements of unpolarized deeply inelastic scattering 
(DIS) of leptons off nuclear targets, unpolarized parton (quarks and gluons) distributions are appreciably modified by the nuclear 
medium over the entire range of values of  
Bjorken $x_B$~\cite{EMC,Frankfurt:1988nt,Arneodo:1992wf,
Geesaman:1995yd,Piller:1999wx}.
Even stronger medium modifications for the polarized parton distributions 
have been predicted for polarized DIS off
nuclear targets~\cite{Guzey:1999rq,Bissey:2001cw,Cloet:2006bq}.

The pattern of nuclear modifications emerging from DIS off nuclear targets 
(and other processes, such as e.g.~proton-nucleus Drell-Yan scattering) can be briefly 
summarized as follows. At small values of Bjorken $x_B$, $x_B < 0.05$, the ratio 
$F_{2}^A(x,Q^2)/[AF_{2}^N(x,Q^2)] <1$,
 where $F_{2}^A(x,Q^2)$ and $F_{2}^N(x,Q^2)$ are the inclusive nuclear and nucleon
structure functions, respectively. This suppression is called nuclear shadowing
and is explained as the effect of the attenuation due to multiple coherent
interactions with the target nucleons~\cite{Piller:1999wx}.
 The effect of nuclear shadowing increases with the target atomic number $A$
as $A^{1/3}$ and is as large as 30\% for heavy targets. 
As one increases $x_B$, $0.05 < x_B < 0.2$, the ratio
$F_{2}^A(x,Q^2)/[AF_{2}^N(x,Q^2)]$ increases above unity by a few percent. 
This enhancement is called antishadowing. While no widely accepted explanation of
antishadowing exists, it can be dynamically generated by taking into account both
the Pomeron and Reggeon exchanges in the interaction of the virtual photon with the
target nucleons~\cite{Brodsky:1989qz}
as well as by the excess of pions in nuclei~\cite{Ericson:1984vt}.
For intermediate values of $x_B$, 
$0.2 < x_B \leq 0.8$, the ratio $F_{2}^A(x,Q^2)/[AF_{2}^N(x,Q^2)]$
is again less than unity and this is usually what is 
called the EMC effect~\cite{EMC}.
It is important to point out that,
while there is no universal and generally accepted explanation of
the EMC effect, it cannot be explained by traditional nuclear physics, where
the nucleus consists of nucleons whose properties are not modified by the
nuclear environment~\cite{Frankfurt:1988nt,Bickerstaff:1989ch}.
The large number of approaches and models for the explanation of the 
EMC effect can be grouped into two large classes~\cite{Arneodo:1992wf}:
 the models introducing
non-nucleon degrees of freedom 
(such as e.g.~the pion cloud~\cite{Llewellyn Smith:1983qa,Ericson:1983um}) 
and the models
assuming some kind of modifications of the nucleons themselves in the nuclear medium
which mentioned earlier~\cite{Guichon,QMCII,Saito:2005rv,Mineo:2003vc,Miller:2001tg,Smith:2002ci,Smith:2003hu}. Our analysis falls into the latter category.
Finally, in the large $x$ limit ($x_B > 0.8$), the ratio
$F_{2}^A(x,Q^2)/[AF_{2}^N(x,Q^2)]>1$ as a consequence of Fermi motion and the fact
$F_{2}^N(x,Q^2)$ vanishes in the $x_B \to 1$ limit.

It should be noted that pion excess models (some of them are mentioned above) automatically lead to the enhancement of sea quarks in nuclei, which seems to contradict the nuclear Drell-Yan data from FNAL~\cite{Alde:1990im}. 
However, a number of recent theoretical papers challenges the ''naive'' 
relation of the nuclear Drell-Yan rates to the nuclear sea quark parton distributions
by discussing initial-state interactions of the quarks going into the nucleus
that lower the effective momentum of the quark at the point where it
annihilates. This can give very big corrections, see e.g.~\cite{Duan:2008qt,ChunGui:2008gn,Duan:2006hp,Arleo:2002ki}.

There has also been considerable interest in the possible modification 
of the bound nucleon elastic form factors.
The polarization transfer measurement in the
$^4{\rm He}(\vec{e},e^{\prime}\vec{p})^3{\rm H}$ reaction at 
the Hall A Jefferson Lab experiment~\cite{Strauch:2002wu,Malace:2008gf}
probes the possible medium modifications of the bound-nucleon form
factors and can be described either by the inclusion of the 
modified elastic form factors as predicted by the Quark-Meson Coupling (QMC) model~\cite{Lu:1997mu} or by the inclusion of the strong 
charge-exchange final-state interaction (FSI)~\cite{Schiavilla:2004xa}.
However, such a strong FSI may not be consistent with the induced 
polarization data -- see Ref.~\cite{Malace:2008gf} for details.
In addition to the modifications of inclusive structure functions (parton distributions)
and elastic form factors of the bound nucleon, the QMC model~\cite{Guichon,QMCII}  predicts modifications 
of various hadron properties in a nuclear medium~\cite{Saito:2005rv}.

Deeply Virtual Compton scattering (DVCS) interpolates between the inclusive DIS 
and elastic scattering reactions
(see Refs.~\cite{Ji:1998pc,Goeke:2001tz,Diehl:2003ny,Belitsky:2005qn} for reviews). 
Therefore, it is natural to expect that generalized parton distributions (GPDs) of the bound nucleon, which are probed by various observables measured in DVCS, 
should also be modified in the nuclear medium. 
An early 
investigation~\cite{Liuti:2005gi,Liuti:2006dx} of such modifications in DVCS on $^4$He
assumed that in-medium nucleon GPDs are modified through the kinematic off-shell effects 
associated with the modification of the relation between
the struck quark's transverse momentum and its virtuality. 

On the experimental side, DVCS on $^4$He in the coherent (the target nucleus remains
intact) and incoherent (the target nucleus breaks up) regimes will be measured 
at Jefferson Lab~\cite{He4-JLab}. 
The expected experimental accuracy will be sufficiently high 
to distinguish between different theoretical predictions 
and to extract the  effects of the medium modifications of the bound nucleon GPDs. 
Note that the first data on coherent and incoherent DVCS on
a wide range of nuclear targets was taken and analyzed by the Hermes 
collaboration~\cite{Ellinghaus:2002zw}. However, 
the accuracy of the data was not sufficient to extract 
the relatively small effects associated with medium modifications.

In this work, we compute medium modifications of the bound 
proton GPDs and their influence 
on {\it incoherent} DVCS on nuclear targets, 
$e A \to e^{\prime} \gamma p X$, where $A$ denotes any
nuclear target. 
As a practical application, 
we consider the beam-spin DVCS asymmetry, $A_{\rm LU}$,
for a $^4$He nucleus,    
since this DVCS observable will soon be measured at 
Jefferson Lab~\cite{He4-JLab}. 
We find the following trend for the ratio of the 
bound to free proton beam-spin asymmetries:  
$A_{\rm LU}^{p^{\ast}}/A_{\rm LU}^p < 1$ for 
small $t$ and $x_B$, and $A_{\rm LU}^{p^{\ast}}/A_{\rm LU}^p > 1$ as $t$ 
and $x_B$ are increased.
The deviation of $A_{\rm LU}^{p^{\ast}}/A_{\rm LU}^p$
from unity arises mainly from the medium modification of the 
bound proton elastic form factor, $F_2^{p^{\ast}}(t)$. 

% Main part

\begin{figure}[t]
\begin{center}
\epsfig{file=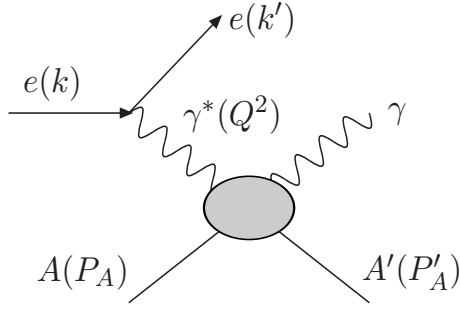,width=6cm,height=4cm}
\caption{Kinematics of DVCS on a nuclear target.}
\label{fig:diagram_dvcs}
\end{center}
\end{figure}

The kinematics of DVCS on a hadronic (nuclear) target, 
$e(k) A(P_A) \to e(k^{\prime}) \gamma A^{\prime}(P_A^{\prime})$, 
is presented in Fig.~\ref{fig:diagram_dvcs}.
The corresponding scattering amplitude reads
\begin{equation}
{\cal T}_{\rm DVCS}^A=-{\bar u}(k^{\prime}) \gamma_{\mu} u(k) \frac{1}{Q^2} H^{\mu \nu} \epsilon_{\nu}^{\ast} \,,
\label{eq:DVCSamplitude}
\end{equation}
where the spinor $u(k)$ [${\bar u}(k^{\prime})$] corresponds to the initial
[final] lepton. $Q^2$ is the virtuality of 
the exchanged photon, and 
$\epsilon_{\nu}^{\ast}$ is the polarization vector of the final real photon.
Note that the final nuclear state $A^{\prime}$ could be both elastic (coherent
DVCS) and inelastic (incoherent DVCS).

Information on the target response is contained in the DVCS hadronic tensor,
$H^{\mu \nu}$, which is defined as a matrix element of the $T$-product of 
two electromagnetic currents,
\begin{equation}
H^{\mu \nu}=-i \int d^4 x\, e^{-i\,q \cdot x} \langle P_A^{\prime}| T\{J^{\mu}(x) J^{\nu}(0)\}|P_A \rangle \,, 
\label{eq:ht_1}
\end{equation}
where $q$ ($-q^2=Q^2$) is the momentum of the virtual photon.
To the leading twist accuracy,
$H^{\mu \nu}$ of a spinless nucleus is expressed in terms of a single
generalized parton distribution, $H^A$, convoluted with the hard scattering
coefficient function $C^{+}(x,\xi_A)$, see e.g.~Ref.~\cite{Goeke:2001tz},\\
\begin{equation}
H^{\mu \nu}=-g^{\mu \nu}_{\perp} \int^{1}_{-1} dx \,C^{+}(x,\xi_A)
H^A(x,\xi_A,t,Q^2) \equiv -g^{\mu \nu}_{\perp} \,{\cal H}^A(\xi_A,t,Q^2) \,,
\label{eq:ht_2}
\end{equation}
where $g^{\mu \nu}_{\perp}=g^{\mu \nu}-\tilde{p}^{\mu} n^{\nu}-\tilde{p}^{\nu} n^{\mu}$
is defined by the two light-like
vectors $\tilde{p}=1/\sqrt{2}(1,0,0,1)$ and $n=1/\sqrt{2}(1,0,0,-1)$, and 
$C^{+}(x,\xi_A)=1/(x-\xi_A+i \epsilon)+1/(x+\xi_A-i \epsilon)$.
The function ${\cal H}^A$ is often called the Compton form factor (CFF). 
It depends on the momentum transfer,
$t=(P_A^{\prime}-P_A)^2$, the longitudinal momentum transfer (skewedness),\\
$\xi_A=-(P_A^{\prime}-P_A) \cdot n/(P_A^{\prime}+P_A) \cdot n \approx x_A/(2-x_A)$, 
where $x_A=Q^2/(2 P_A \cdot q)$ is the Bjorken variable,
and the virtuality, 
$Q^2$.

\begin{figure}[t]
\begin{center}
\epsfig{file=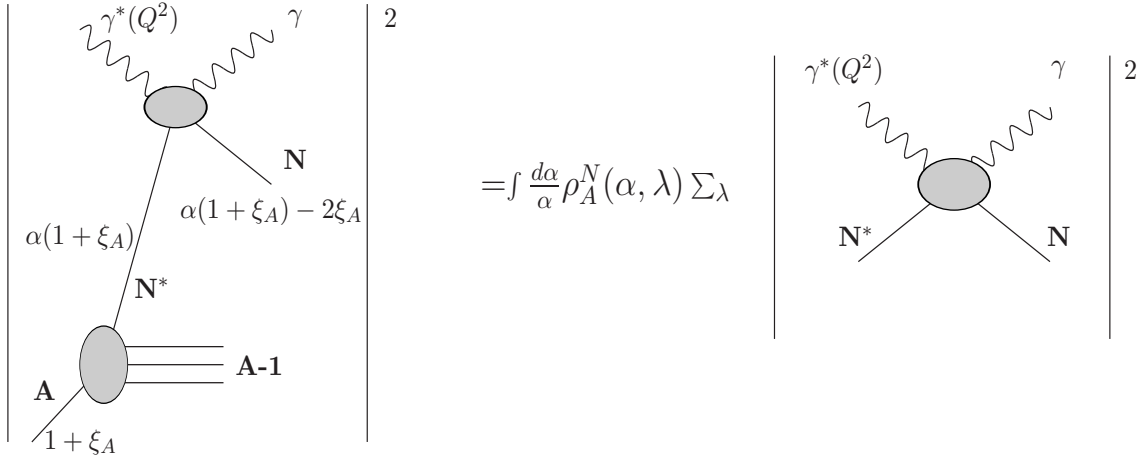,width=15cm,height=6cm}
\caption{Illustration of the connection between incoherent 
DVCS on a nuclear target and DVCS on a bound nucleon. 
(See Eq.~(\ref{eq:connection_main})). The
relevant light-cone fractions are also indicated.}
\label{fig:diagram_mm}
\end{center}
\end{figure}

Incoherent DVCS on a nuclear target occurs on a single nucleon,
$e A \to e^{\prime} \gamma N X$. 
Therefore, the squared amplitude of DVCS on a nuclear target,
$|{\cal T}_{\rm DVCS}^A|^2$, can be expressed in terms 
of the squared amplitude of DVCS on 
the bound nucleons, 
$|{\cal T}_{\rm DVCS}^{N^{\ast}}|^2$. This is graphically presented in 
Fig.~\ref{fig:diagram_mm}. 
Below we give the derivation and
explain the notation in Fig.~\ref{fig:diagram_mm}.
In our work, we follow the example of the derivation of GPDs of deuterium by Cano and
Pire~\cite{Cano:2003ju}. For a similar formalism,
see also~\cite{Liuti:2005gi,Liuti:2006dx}.

The connection between $|{\cal T}_{\rm DVCS}^A|^2$
and $|{\cal T}_{\rm DVCS}^{N^{\ast}}|^2$ 
can be derived 
straightforwardly
using the notion of the nuclear light-cone (LC) wave function.
In the formalism of LC quantization, each state is characterized by its
plus-momentum, $p^+=p \cdot n=(p^0+p^3)/\sqrt{2}$, the transverse momentum,
 $\vec{p}_{\perp}$,
and the helicity, $\lambda$. 
The nuclear state $|P_A \rangle$ 
is expressed in terms of the nuclear
LC wave function, $\phi_A$, and the product of nucleon states, 
$|p_i^+,\vec{p}_{\perp i},\lambda_i\rangle$~\cite{Lepage:1980fj},
\begin{eqnarray}
|P_A^+,\vec{P}_{\perp A} \rangle &=& \sum_{\lambda_i} \int \prod_{i=1}^A \frac{d \alpha_i}{\sqrt{\alpha_i}}
\frac{d^2 \vec{k}_{\perp i}}{16 \pi^3} 16 \pi^3 \delta\left(\sum_{j=1}^A \alpha_j-1\right)
\delta\left(\sum_{j=1}^A \vec{k}_{\perp j}\right) \nonumber\\
&\times& \phi_A(\alpha_i,\vec{k}_{\perp i},\lambda_i)|\alpha_iP_A^+, \vec{k}_{\perp i}
+\alpha_i \vec{P}_{\perp A},\lambda_i\rangle \,,
\label{eq:lc_wf}
\end{eqnarray}
where $\alpha_i=p_i^+/P_A^+$ is the fraction of the 
nucleus plus-momentum carried by nucleon~$i$.

Substituting Eq.~(\ref{eq:lc_wf}) for the initial nuclear state 
$|P_A \rangle$ in the nuclear hadronic tensor defined by 
Eq.~(\ref{eq:ht_1}), and using the assumption
that the final nuclear state $|P_A^{\prime} \rangle$
consists of an active nucleon $N^{\ast}$  and $A-1$ spectators (see 
Fig.~\ref{fig:diagram_mm}), one  
obtains the relation between $|{\cal T}_{\rm DVCS}^A|^2$
and $|{\cal T}_{\rm DVCS}^{N^{\ast}}|^2$,
\begin{equation}
|{\cal T}_{\rm DVCS}^A(\xi_A,t)|^2=\int^1_{\alpha_{\rm min}} \frac{d \alpha}{\alpha} \rho_A^N(\alpha,\lambda)
 \sum_{\lambda}|{\cal T}_{\rm DVCS}^{N^{\ast}}(\xi_N,t)|^2 \,,
\label{eq:connection_main}
\end{equation}
where the nucleon light-cone distribution $\rho_A^N(\alpha,\lambda)$ 
is defined in terms of the nuclear LC wave function~\cite{Frankfurt:1981mk}:
\begin{eqnarray}
\rho_A^N(\alpha,\lambda)&=&\int \frac{d^2 \vec{k}_{\perp}}{16 \pi^3}
\sum_{\lambda_i}
\prod_{i=2}^A \frac{d \alpha_i \,d^2 \vec{k}_{\perp i}}{16 \pi^3}\,
\delta\left(\alpha+\sum_{j=2}^A \alpha_j-1\right)
16 \pi^3 \delta\left(\vec{k}_{\perp}+\sum_{j=2}^A \vec{k}_{\perp j}\right)
\nonumber\\
&\times&
|\phi_A(\alpha,\vec{k}_{\perp},\lambda,\alpha_i,\vec{k}_{\perp i},\lambda_i)|^2
\,.
\label{eq:rho}
\end{eqnarray}
Here the light-cone fraction $\alpha$ and the transverse momentum
$\vec{k}_{\perp}$ refer to the interacting nucleon, while $\alpha_i$ and 
$\vec{k}_{\perp i}$ refer to the spectator nucleons.
The distribution $\rho_A^N(\alpha,\lambda)$ introduced above is normalized to unity, 
\begin{equation}
\sum_{\lambda} \int_0^1 d \alpha \, \rho_A^N(\alpha,\lambda)=1 \,.
\label{eq:normalization}
\end{equation} 

In Fig.~\ref{fig:diagram_mm}, we also show the relevant light-cone momentum 
fractions
(we use the standard symmetric frame~\cite{Ji:1998pc}):
the nucleus carries the plus-momentum $P_A^+=(1+\xi_A) {\bar P}_A^{+}$,
where ${\bar P}_A^+=(P_A^{+}+P_A^{\prime +})/2$;
the active nucleon has $p^+=\alpha(1+\xi_A) {\bar P}_A^{+}$ in the initial state and   
$p^{\prime +}=(\alpha(1+\xi_A)-2 \xi_A){\bar P}_A^{+}$ in the 
final state.
The requirement $p^{\prime +} \geq 0$
 determines the minimal value of $\alpha$, $\alpha_{\rm min}=2\xi_A/(1+\xi_A)$.

In Eq.~(\ref{eq:connection_main}),
the skewedness, $\xi_N$, 
is defined with respect to the active nucleon in
the symmetric frame~\cite{Cano:2003ju}:
\begin{equation}
\xi_N=\frac{\xi_A}{\alpha(1+\xi_A)-\xi_A} \,.
\label{eq:xiN}
\end {equation} 

The light-cone distribution $\rho_A^N(\alpha)$ is peaked around 
$\alpha \approx 1/A$. From our experience with the EMC effect~\cite{EMC},
it is well known that, except for large $x_B$,
 the effect of Fermi motion is small~\cite{Frankfurt:1988nt,Arneodo:1992wf,Geesaman:1995yd}.
 Therefore, we neglect Fermi motion of the bound nucleon and 
evaluate $|{\cal T}_{\rm DVCS}^{N^{\ast}}(\xi_N,t)|^2$ 
at $\alpha=1/A$. Using
the normalization condition of Eq.~(\ref{eq:normalization}), 
we obtain an approximate expression for $|{\cal T}_{\rm DVCS}^A|^2$,\\
\begin{equation}
|{\cal T}_{\rm DVCS}^A(\xi_A,t)|^2\simeq\sum_N
\frac{1}{2} \sum_{\lambda}|{\cal T}_{\rm DVCS}^{N^{\ast}}(\langle \xi_N \rangle,t)|^2 \,,
\label{eq:connection_main_2}
\end{equation}
where the factor $1/2$ comes from the normalization 
condition of Eq.~(\ref{eq:normalization}), and
the average nucleon skewedness, $\langle \xi_N \rangle$, is defined as  
\begin{equation}
\langle \xi_N \rangle  \equiv \frac{\xi_A}{\frac{1}{A}(1+\xi_A)-\xi_A} \,.
\label{eq:xiA}
\end {equation}

To compare with the experiment, it is convenient to
rescale $x_A$ and to define the Bjorken variable, $x_B$, with respect to the nucleon:
\begin{equation}
x_B \equiv A \,\frac{Q^2}{2 P_A \cdot q} \equiv A\,x_A
\,.
\label{eq:xA}
\end{equation}

The corresponding skewedness $\xi$, $\xi=x_B/(2-x_B)$, coincides with
that given by Eq.~(\ref{eq:xiA}).  
Using the Bjorken $x_B$ of Eq.~(\ref{eq:xA}), 
and the fact that both sides of
Eq.~(\ref{eq:connection_main_2}) depend on the same skewedness $\xi$, 
we obtain:
\begin{equation}
|{\cal T}_{\rm DVCS}^A(\xi,t)|^2=\sum_N
\frac{1}{2} \sum_{\lambda}|{\cal T}_{\rm DVCS}^{N^{\ast}}(\xi,t)|^2 \,.
\label{eq:connection_main_3}
\end {equation}
The interpretation of  Eq.~(\ref{eq:connection_main_3}) is 
intuitive: the probability of incoherent DVCS on a nuclear target is a sum of 
the probabilities of DVCS on individual nucleons.

\begin{figure}[t]
\begin{center}
\epsfig{file=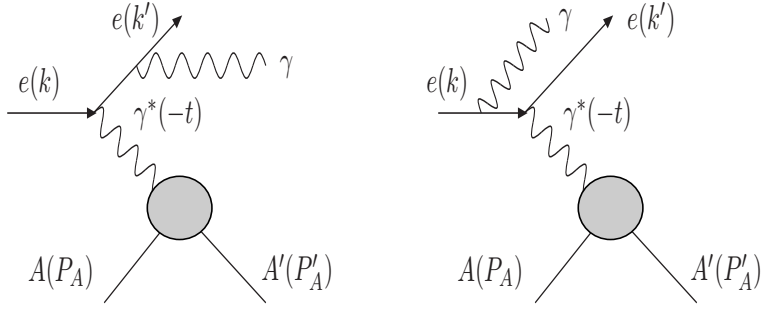,width=10cm,height=4cm}
\caption{The Bethe-Heitler process.}
\label{fig:diagram_bh}
\end{center}
\end{figure}

Since Eq.~(\ref{eq:connection_main_3}) is based on 
the decomposition of Eq.~(\ref{eq:lc_wf}) and does not depend on 
the type of the elementary
interaction with the active nucleon, similar relations also hold for 
the Bethe-Heitler (BH) amplitude (see Fig.~\ref{fig:diagram_bh}) and 
for the interference between the DVCS and BH amplitudes 
(see Ref.~\cite{Belitsky:2001ns} for details of the definitions of the BH and interference
amplitudes): 
\begin{eqnarray}
|{\cal T}_{\rm BH}^A(\xi,t)|^2&=&\sum_N
\frac{1}{2} \sum_{\lambda}|{\cal T}_{\rm BH}^{N^{\ast}}(\xi,t)|^2 \,,
\nonumber\\
|{\cal I}^A(\xi,t)|^2&=&\sum_N
\frac{1}{2} \sum_{\lambda}|{\cal I}^{N^{\ast}}(\xi,t)|^2 \,.
\label{eq:connection_main_4}
\end{eqnarray}

The practical corollary of Eqs.~(\ref{eq:connection_main_3}) and (\ref{eq:connection_main_4}) is the following: expressions for  
DVCS observables (cross section asymmetries) in incoherent nuclear DVCS on a
spinless nuclear target are exactly the same as those for the sum of 
individual bound nucleons. 

In this work, we apply Eqs.~(\ref{eq:connection_main_3}) 
and (\ref{eq:connection_main_4}) to  incoherent DVCS on 
$^4$He in the situation when a proton in the final state is detected,
$e ^4{\rm He} \to e^{\prime} \gamma p X$. In this case, the neutrons do not
contribute and  Eqs.~(\ref{eq:connection_main_3}) and (\ref{eq:connection_main_4}) become 
\begin{eqnarray}
|{\cal T}_{\rm DVCS}^{A}(\xi,t)|^2 &=& \sum_{\lambda}|{\cal T}_{\rm DVCS}^{p^{\ast}}(\xi,t)|^2 \,, \nonumber\\
|{\cal T}_{\rm BH}^A(\xi,t)|^2 &=& \sum_{\lambda}|{\cal T}_{\rm BH}^{p^{\ast}}(\xi,t)|^2 \,,
\nonumber\\
|{\cal I}^A(\xi,t)|^2&=&\sum_{\lambda}|{\cal I}^{p^{\ast}}(\xi,t)|^2 \,.
\label{eq:connection_main_5}
\end{eqnarray}
Note that although Eq.~(\ref{eq:connection_main_5}) does not contain 
an explicit reference to the Fermi motion of the bound nucleon,
it does implicitly contain some effects of the Fermi motion through
the self-consistent change of the internal  structure of the 
bound nucleon via the 
medium-modified proton elastic form factors (see below).

As we mentioned in the Introduction, the GPDs of the bound nucleon 
may generally differ from the GPDs of the free nucleon. 
In this work, we assume that the GPDs of the bound proton  
are modified in proportion to the corresponding
bound proton elastic form factors,
\begin{eqnarray}
H^{q/p^{\ast}}(x,\xi,t,Q^2)&=&\frac{F_1^{p^{\ast}}(t)}{F_1^p(t)} \,
H^{q}(x,\xi,t,Q^2) \,, \nonumber\\
E^{q/p^{\ast}}(x,\xi,t,Q^2)&=&\frac{F_2^{p^{\ast}}(t)}{F_2^p(t)} 
E^{q}(x,\xi,t,Q^2) \,, \nonumber\\
\tilde{H}^{q/p^{\ast}}(x,\xi,t,Q^2)&=&\frac{G_1^{\ast}(t)}{G_1(t)} \,
\tilde{H}^{q}(x,\xi,t,Q^2) \,,
\label{eq:gpds_mm}
\end{eqnarray}
where the GPDs $H^{q/p^{\ast}}$, $E^{q/p^{\ast}}$ and $\tilde{H}^{q/p^{\ast}}$
and the elastic form factors $F_1^{p^{\ast}}$ (Dirac form factor), 
$F_2^{p^{\ast}}$ (Pauli form factor) and $G_1^{\ast}$ (axial form factor)
 refer to the bound proton, while
$H^{q}$, $E^{q}$, and $\tilde{H}^{q}$ and $F_1^{p}$, $F_2^{p}$ and $G_1$
 refer to those of the free proton. 
The assumption of Eq.~(\ref{eq:gpds_mm}) is rather simple, since 
the medium-modifications result only in the $t$-dependent
 renormalization and do not change the shape of the in-medium GPDs. 
The GPDs $H^{q/p^{\ast}}(x,\xi,t,Q^2)$ and 
$E^{q/p^{\ast}}(x,\xi,t,Q^2)$ in a $^4$He nucleus 
are constrained to reproduce 
the extracted bound proton elastic electromagnetic form factors 
after integration over $x$, as the QMC model 
predicted~\cite{Strauch:2002wu} (see below). 
Note also that we have ignored the insignificant kinematically-suppressed contribution of 
the GPD $\tilde{E}$ to the DVCS beam-spin asymmetry~\cite{Belitsky:2001ns}.

\begin{figure}[t]
\begin{center}
\epsfig{file=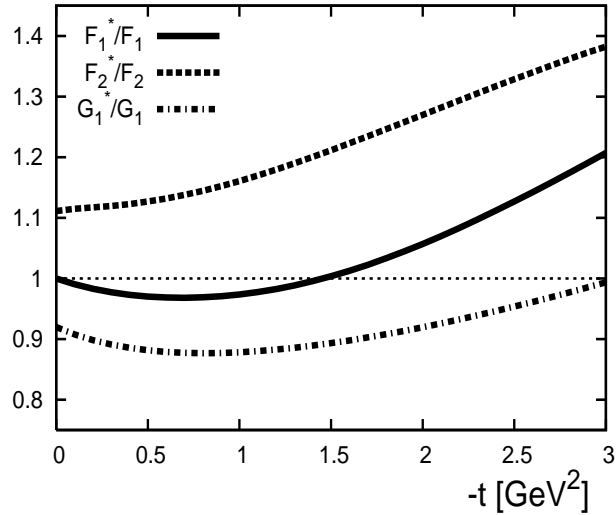,width=9cm,height=7cm}
\caption{The bound ($^4$He) to free proton ratios of elastic
form factors $F_1^{p^{\ast}}(t)/F_1^{p}(t)$, $F_2^{p^{\ast}}(t)/F_2^{p}(t)$ 
and $G_1^{\ast}(t)/G_1(t)$ as functions of the momentum transfer $t$,
see Eq.~(\ref{eq:density_average}).
}
\label{fig:He4_MM2}
\end{center}
\end{figure}

The bound proton form factors have been calculated in the 
QMC model~\cite{Lu:1997mu,QMCgA,QMCnuA}. 
Since these form factors
depend on the nuclear density, the in-medium form factors in 
Eq.~(\ref{eq:gpds_mm}) must be averaged over 
the nuclear density distribution in $^4$He ($A=^4$He below),
\begin{eqnarray}
F_1^{p^{\ast}}(t)&=&\int d^3 \vec{r} \,\rho_A(r)\, 
F_1^{p^{\ast}}(t,\rho_A(r)) \,, 
\nonumber\\
F_2^{p^{\ast}}(t)&=&\int d^3 \vec{r} \,\rho_A(r)\, 
F_2^{p^{\ast}}(t,\rho_A(r)) \,,
\nonumber\\
G_1^{\ast}(t)&=&\int d^3 \vec{r} \,\rho_A(r)\, 
G_1^{\ast}(t,\rho_A(r)) \,,
\label{eq:density_average}
\end{eqnarray}
where $F_1^{p^{\ast}}(t,\rho_A(r))$, $F_2^{p^{\ast}}(t,\rho_A(r))$
and $G_1^{\ast}(t,\rho_A(r))$ are the nuclear density-dependent
bound proton form factors, and $\rho_A(r)$ ($ \equiv \rho_{^4{\rm He}}(r)$)  
is the nuclear density distribution in $^4$He 
calculated in Ref.~\cite{Saito:1997ae}.
In Fig.~\ref{fig:He4_MM2}, we show the resulting ratios
$F_1^{p^{\ast}}(t)/F_1^{p}(t)$, $F_2^{p^{\ast}}(t)/F_2^{p}(t)$ 
and $G_1^{\ast}(t)/G_1(t)$ 
as functions of $-t$~\cite{Lu:1997mu,QMCgA,QMCnuA}.

For the free proton GPDs, we use the double distribution 
model~\cite{Radyushkin:1998es} based on valence quark PDFs.
In particular, we use 
\begin{eqnarray}
H^q(x,\xi,t,Q^2)&=&\int^{1}_{0} d \beta \int^{1-|\beta|}_{-1+|\beta|}
d \alpha\, \delta(\beta+\alpha\, \xi-x) \pi(\beta,\alpha) \beta^{-\alpha^{\prime}(1-\beta)t}q_v(\beta,Q^2) \,,
\nonumber\\
E^q(x,\xi,t,Q^2)&=&\int^{1}_{0} d \beta \int^{1-|\beta|}_{-1+|\beta|}
d \alpha\, \delta(\beta+\alpha\, \xi-x) \pi(\beta,\alpha)\beta^{-\alpha^{\prime}(1-\beta)t} e_v^q(\beta,Q^2) \,,
\nonumber\\
\tilde{H}^q(x,\xi,t,Q^2)&=&\int^{1}_{0} d \beta \int^{1-|\beta|}_{-1+|\beta|}
d \alpha\, \delta(\beta+\alpha\, \xi-x) \pi(\beta,\alpha) \beta^{-\alpha^{\prime}(1-\beta)t}\Delta q_v(\beta,Q^2) \,,
\label{eq:dd}
\end{eqnarray}
where $q_v$ and $\Delta q_v$ are the valence unpolarized and polarized 
quark PDFs, respectively, while $e_v^q(\beta)$ is the valence part of the 
forward limit of the GPD $E^q$.
The profile function $\pi(\beta,\alpha)$ is taken 
in a standard form~\cite{Goeke:2001tz}:
\begin{equation}
\pi(\beta,\alpha)=\frac{3}{4}\frac{(1-\beta)^2-\alpha^2}{(1-\beta)^3} \,.
\label{eq:profile}
\end{equation}
The $t$-dependence of GPDs is introduced through 
the Regge theory-motivated factor with the slope parameter 
$\alpha^{\prime}=1.105$ GeV$^{-2}$, which leads to a good description
of the proton and neutron elastic form factors~\cite{Guidal:2004nd}.

For the unpolarized quark PDFs, we 
use the leading-order (LO) CTEQ5L
parameterization~\cite{Lai:1999wy}, while for the polarized quark PDFs, we 
use the LO GRSV 2000 parameterization~\cite{Gluck:2000dy}. 
The model for the forward limit of the GPD $E^q$ is taken from 
Ref.~\cite{Guidal:2004nd}. 
Explicitly, it is given by 
\begin{eqnarray}
e_v^u(x,Q^2) & =& \frac{k_u}{N_u}\,(1-x)^{\eta_u} u_v(x,Q^2) \,,
\nonumber\\
e_v^d(x,Q^2) & =& \frac{k_d}{N_d}\,(1-x)^{\eta_d} d_v(x,Q^2) \,,
\label{eq:model_e}
\end{eqnarray}
where $k_u=1.673$ and $k_d=-2.033$ are the anomalous magnetic moments;
$\eta_u=1.713$ and $\eta_d=0.566$ are determined from fits to the nucleon
elastic form factors; $N_u$ and $N_d$ are the normalization factors,
\begin{eqnarray}
N_u &=& \int^1_0 dx\, (1-x)^{\eta_u} u_v(x,Q^2) \,,
\nonumber\\
N_d &=& \int^1_0 dx\, (1-x)^{\eta_d} d_v(x,Q^2) \,.
\label{eq:normal_e}
\end{eqnarray}

\begin{figure}[t]
\begin{center}
\epsfig{file=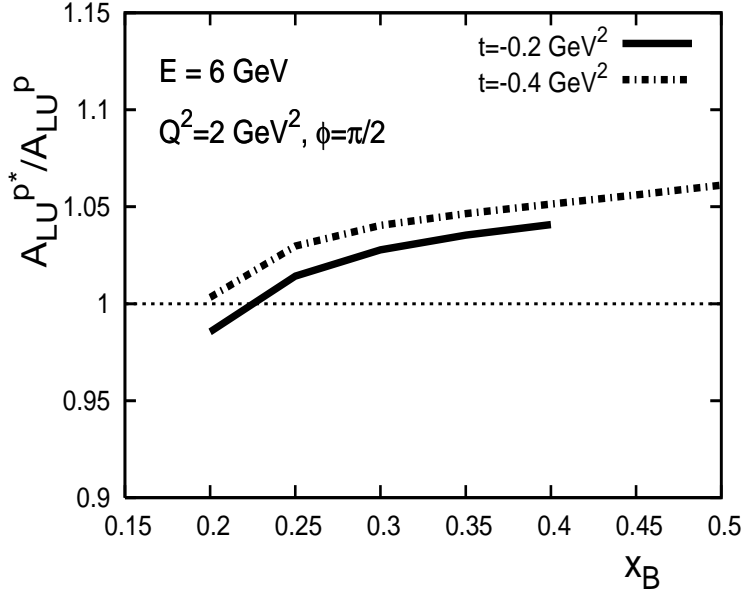,width=10cm,height=8cm}
\caption{The ratio of the bound 
(incoherent $^4$He) to free proton beam-spin DVCS asymmetries,
$A_{\rm LU}^{p^{\ast}}(\phi)/A_{\rm LU}^{p}(\phi)$, as a function of 
Bjorken $x_B$ at $E=6$ GeV, $Q^2=2$ GeV$^2$, $\phi=\pi/2$ and two 
values of $t$.}
\label{fig:He4_Beam_Spin_MM_DD_xBdep}
\end{center}
\end{figure}

\begin{figure}[t]
\begin{center}
\epsfig{file=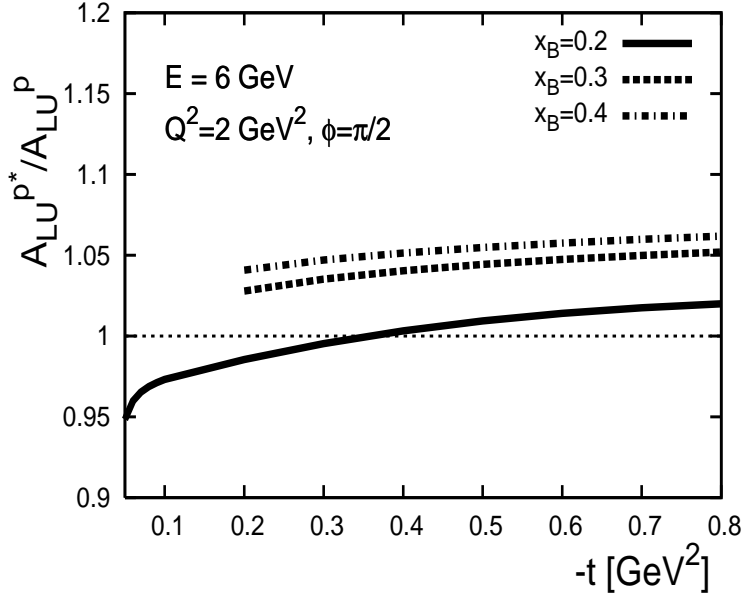,width=10cm,height=8cm}
\caption{The ratio of the bound (incoherent $^4$He) to 
free proton beam-spin DVCS asymmetries,
$A_{\rm LU}^{p^{\ast}}(\phi)/A_{\rm LU}^{p}(\phi)$, as a function of 
the momentum transfer $t$ at $E=6$ GeV, $Q^2=2$ GeV$^2$, $\phi=\pi/2$
and three values of $x_B$.}
\label{fig:He4_Beam_Spin_MM_DD_tdep}
\end{center}
\end{figure}

In summary, the bound proton GPDs are given by 
Eqs.~(\ref{eq:dd})--(\ref{eq:normal_e}). 
Since for the case of incoherent DVCS on $^4$He, 
$e ^4{\rm He} \to e^{\prime} \gamma p X$, the scattering amplitudes squared 
are the same as those for the bound proton  
(see Eq.~(\ref{eq:connection_main_5})), 
we may use the standard formalism developed for 
the free nucleon~\cite{Belitsky:2001ns} 
to calculate various DVCS observables (cross section asymmetries).
Our results are presented in Figs.~\ref{fig:He4_Beam_Spin_MM_DD_xBdep}
and \ref{fig:He4_Beam_Spin_MM_DD_tdep}.

In Fig.~\ref{fig:He4_Beam_Spin_MM_DD_xBdep} we present the ratio of 
the bound (incoherent $^4$He) to free proton beam-spin DVCS asymmetries,
$A_{\rm LU}^{p^{\ast}}(\phi)/A_{\rm LU}^{p}(\phi)$, as a function of 
Bjorken $x_B$ at the fixed energy of the lepton beam, $E=6$ GeV, and  
virtuality $Q^2=2$ GeV$^2$. This asymmetry is measured with a linearly
polarized lepton beam and an unpolarized target. 
The $A_{\rm LU}(\phi)$ asymmetry 
is mostly sensitive to the imaginary part of the Compton form factor, 
${\rm Im}{\cal H}^A$ (see Eq.~(\ref{eq:ht_2})), and behaves as 
$A_{\rm LU} \propto {\rm Im}{\cal H}^A \sin \phi$, where $\phi$ is the angle
between the leptonic and hadronic (production) planes. 
(See Ref.~\cite{Belitsky:2001ns} for the details.)
Note that
in Fig.~\ref{fig:He4_Beam_Spin_MM_DD_xBdep}, 
$A_{\rm LU}(\phi)$ is evaluated at $\phi=\pi/2$.

As seen from Fig.~\ref{fig:He4_Beam_Spin_MM_DD_xBdep}, effects of
the medium-modifications in the kinematic region under study do not exceed 
$\sim$6\%.
Their trend can be understood by analyzing 
the approximate expression for 
$A_{\rm LU}(\phi)$~\cite{Belitsky:2001ns}, 
\begin{equation}
A_{\rm LU}(\phi)  \propto   {\rm Im} \left(F_1^{p^{\ast}}{\cal H}^{p^{\ast}}+\frac{x_B}{2-x_B} \left(F_1^{p^{\ast}}+F_2^{p^{\ast}}\right)
\tilde{{\cal H}}^{p^{\ast}}-\frac{t}{4 m_N^2} F_2^{p^{\ast}}{\cal E}^{p^{\ast}} 
\right) 
 \Big/  f(F_1^{p^{\ast}},F_2^{p^{\ast}}) \sin \phi \,,
\label{eq:alu}
\end{equation}
where ${\cal H}^{p^{\ast}}$, ${\cal E}^{p^{\ast}}$ and 
$\tilde{{\cal H}}^{p^{\ast}}$ are the Compton form factors of the respective
bound nucleon GPDs; 
$f(F_1^{p^{\ast}},F_2^{p^{\ast}})$ is a certain
function (dominated by the Bethe-Heitler amplitude squared) of the elastic form 
factors.
Note that the argument of the elastic form
factors is the invariant momentum transfer $t$ 
(see Fig.~\ref{fig:diagram_bh}).

At small $x_B$ and $t$, the contributions of  $\tilde{{\cal H}}^{p^{\ast}}$ and
${\cal E}^{p^{\ast}}$ in Eq.~(\ref{eq:alu}) are unimportant and  
$A_{\rm LU}^{p^{\ast}}(\phi)/A_{\rm LU}^{p}(\phi)<1$ 
because of  the increase of  $f(F_1^{p^{\ast}},F_2^{p^{\ast}})$ for the 
bound nucleon.  
This comes mainly  from the enhancement, 
$F_2^{p^{\ast}} > F_2^{p}$, in $^4$He. 
(See Fig.~\ref{fig:He4_MM2}.) 

As $x_B$ and $t$ are increased,
$\tilde{{\cal H}}^{p^{\ast}}$ and ${\cal E}^{p^{\ast}}$ in Eq.~(\ref{eq:alu}) 
start to play a progressively more
important role (the contribution of  $\tilde{{\cal H}}^{p^{\ast}}$ 
is more important). Thus, the medium-enhancement  of the term
proportional to $(F_1^{p^{\ast}}+F_2^{p^{\ast}})\tilde{{\cal H}}^{p^{\ast}}$
wins over the enhancement of the denominator in Eq.~(\ref{eq:alu}), and
makes $A_{\rm LU}^{p^{\ast}}(\phi)/A_{\rm LU}^{p}(\phi) > 1$.

In Fig.~\ref{fig:He4_Beam_Spin_MM_DD_tdep} we present the ratio
$A_{\rm LU}^{p^{\ast}}(\phi)/A_{\rm LU}^{p}(\phi)$
as a function of the invariant momentum transfer $t$, in the same kinematics as 
in Fig.~\ref{fig:He4_Beam_Spin_MM_DD_xBdep}. 
The size of the medium-modification is
similar to that shown in Fig.~\ref{fig:He4_Beam_Spin_MM_DD_xBdep}, and the trend
of the medium modifications of the ratio 
$A_{\rm LU}^{p^{\ast}}(\phi)/A_{\rm LU}^{p}(\phi)$ 
has a similar interpretation.

Our numerical predictions are based on the particular model of
the nucleon GPDs summarized by Eqs.~(\ref{eq:dd})--(\ref{eq:normal_e}).
We have explicitly checked that taking a different profile function
$\pi(\beta,\alpha)$ in Eq.~(\ref{eq:profile}), 
e.g.~$\pi(\beta,\alpha)=15/16[(1-\beta)^2-\alpha^2]^2/(1-\beta)^5$,
does not change our numerical prediction. 
Since we present our results in the form of the 
ratio of the nuclear to nucleon DVCS asymmetries, 
we expect that 
 details and subtleties of the nucleon GPDs
 should mostly cancel in the ratio and, thus, our predictions summarized in Fig.~\ref{fig:He4_Beam_Spin_MM_DD_xBdep}
and \ref{fig:He4_Beam_Spin_MM_DD_tdep} should be stable against variations of the
parameterization of the nucleon GPDs.

In our analysis, we did not address the issue of possible final state interactions (FSI)
between the produced proton (nucleon) and the remaining $A=3$ system.
In principle, this is a separate, rather involved analysis. However, 
based on the observation that the 
non-charge-exchange FSI for the $^4{\rm He}(\vec{e},e^{\prime}\vec{p})^3{\rm H}$ reaction
are rather small~\cite{Schiavilla:2004xa}
and on the observation that the 
large charge-exchange final-state interaction (FSI) for the same reaction are inconsistent
with the polarization transfer data~\cite{Malace:2008gf}, 
one should not expect FSI for our case of incoherent DVCS, $^4{\rm He}(e,e^{\prime}\gamma p)X$,
that are larger than a few percent. Therefore, 
the theoretical uncertainty associated with the FSI is not large and should not affect 
our conclusions.
%vg December 2008
One should emphasize that the medium modifications of the bound nucleon GPDs and
FSI are two separate effects. Once the effect of FSI for incoherent DVCS on $^4$He is 
estimated, it should be added on the top of the medium modification effects
discussed in the present Letter.

Finally, we would like to compare our results in 
Figs.~\ref{fig:He4_Beam_Spin_MM_DD_xBdep} 
and \ref{fig:He4_Beam_Spin_MM_DD_tdep} with 
the predictions of Liuti and Taneja~\cite{Liuti:2005gi}.
While in our model of the bound proton GPDs in $^4$He, the effects of
Fermi motion, off-shellness, and the internal structure 
change of the bound nucleon are encoded in
the medium-modified proton elastic form factors, the approach of 
Ref.~\cite{Liuti:2005gi} explicitly takes into account such effects 
in the bound nucleon GPD.
Furthermore, the bound nucleon GPDs in the approach of Ref.~\cite{Liuti:2005gi} 
are modified through the kinematic off-shell effects 
associated with the modification of the relation between
the struck quark's transverse momentum and its virtuality. 

First we discuss the $t$-dependence. 
While our prediction for the $t$-dependence of 
$A_{\rm LU}^{p^{\ast}}(\phi)/A_{\rm LU}^{p}(\phi)$ is 
similar to that of Ref.~\cite{Liuti:2005gi}, the size of the nuclear modifications 
is significantly smaller in our case. 
Although the $x_B$-dependence of incoherent DVCS was 
not presented in Ref.~\cite{Liuti:2005gi}, 
the $x_B$-dependence of $A_{\rm LU}^{p^{\ast}}(\phi)/A_{\rm LU}^{p}(\phi)$, 
which is based on the same model as presented in Ref.~\cite{Liuti:2005gi}, 
was given in the proposal of the future Jefferson Lab
experiment~\cite{He4-JLab}.
Our predictions for the $x_B$-dependence of 
$A_{\rm LU}^{p^{\ast}}(\phi)/A_{\rm LU}^{p}(\phi)$ 
are very different both in shape and in size 
from those presented in Ref.~\cite{He4-JLab}. 
In particular, our prediction for the in-medium modification is much smaller 
in magnitude. 
The future Jefferson Lab experiment on DVCS on $^4$He will be able to 
distinguish between our predictions and those of Ref.~\cite{Liuti:2005gi}.

In conclusion, we have studied incoherent DVCS on $^4$He in the 
$^4{\rm He}(e,e^{\prime}\gamma p)X$ reaction, 
which probes medium-modifications
of the bound proton GPDs and elastic form factors.
Assuming that the proton GPDs are modified in proportion to the
corresponding bound proton elastic form factors, as predicted in the Quark-Meson Coupling model, 
we have developed an approach to 
calculate various incoherent nuclear DVCS observables. 
As an example, we have computed the beam-spin  DVCS asymmetry and made predictions for 
the $x_B$- and $t$-dependence of the ratio of the  bound to free proton asymmetries,
$A_{\rm LU}^{p^{\ast}}(\phi)/A_{\rm LU}^{p}(\phi)$. 
We have found that the deviation of 
$A_{\rm LU}^{p^{\ast}}(\phi)/A_{\rm LU}^{p}(\phi)$ from unity 
is as much as $\sim$6\%. 
We checked that our predictions are stable against the variation of the model of the nucleon
GPDs. Also, based on the studies of final state interactions in $^4{\rm He}(\vec{e},e^{\prime}\vec{p})^3{\rm H}$
quasi-elastic scattering, we argue that the effect of the FSI should not exceed a few percent
in our case of incoherent DVCS on $^4$He. 
 
\noindent
{\it Acknowledgments}\\
{\bf Notice}: Authored by Jefferson Science Associates, LLC under U.S. DOE Contract No. DE-AC05-06OR23177. The U.S. Government retains a non-exclusive, paid-up, irrevocable, world-wide license to publish or reproduce this manuscript for U.S. Government purposes.

\end{document}